\documentclass[twocolumn,showpacs]{revtex4}
\usepackage{amsmath}
\usepackage{times}
\usepackage{epsfig}
\usepackage{epstopdf}
\usepackage{graphicx}

\begin{document}

\draft

\title{Solving the accuracy-diversity dilemma via directed random walks}

\author{Jian-Guo Liu}\thanks{jianguo.liu@sbs.ox.ac.uk}
\affiliation{Research Center of Complex Systems Science, University
of Shanghai for Science and Technology, \\ Shanghai 200093,
Peoples's Republic of China} \affiliation{CABDyN Complexity Center,
Sa\"{i}d Business School, University of Oxford, Park End Street,
Oxford, OX1 1HP, United Kingdom}

\author{Kerui Shi}
\affiliation{Research Center of Complex Systems Science, University
of Shanghai for Science and Technology, \\ Shanghai 200093,
Peoples's Republic of China}

\author{Qiang Guo}
\affiliation{Research Center of Complex Systems Science, University
of Shanghai for Science and Technology, \\ Shanghai 200093,
Peoples's Republic of China} \affiliation{Department of Physics,
University of Fribourg, Chemin du Mus\'{e}e 3, CH-1700 Fribourg,
Switzerland}

\begin{abstract} \textnormal{\small {Random walks have been successfully used to measure user or object
similarities in collaborative filtering (CF) recommender systems,
which is of high accuracy but low diversity. A key challenge of CF
system is that the reliably accurate results are obtained with the
help of peers' recommendation, but the most useful individual
recommendations are hard to be found among diverse niche objects. In
this paper we investigate the direction effect of the random walk on
user similarity measurements and find that the user similarity,
calculated by directed random walks, is reverse to the initial
node's degree. Since the ratio of small-degree users to large-degree
users is very large in real data sets, the large-degree users'
selections are recommended extensively by traditional CF algorithms.
By tuning the user similarity direction from neighbors to the target
user, we introduce a new algorithm specifically to address the
challenge of diversity of CF and show how it can be used to solve
the accuracy-diversity dilemma. Without relying on any
context-specific information, we are able to obtain accurate and
diverse recommendations, which outperforms the state-of-the-art CF
methods. This work suggests that the random walk direction is an
important factor to improve the personalized recommendation
performance.}}
\end{abstract}


\maketitle

\section{Introduction}
Due to the rapidly expanding internet and social network, we are
overloaded by the unlimited information on the World Wide Web
\cite{Faloutsos99}. For instance, one has to choose among thousands
of candidate commodities to shop online, and finds the relevant
information from billions of Web pages. Comprehensive exploration
for each user is infeasible \cite{Broder00}. Consequently, how to
efficiently help people obtain information that they truly need is a
challenging task nowadays \cite{Adomavicius05}. A landmark for
information filtering is the use of search engine, by which users
could find the relevant web pages with the help of properly chosen
keywords \cite{Brin1998,Kleinberg1999}. However, sometimes users'
tastes or preferences evolve with time and can not be accurately
expressed by keywords, and search engines do not take into account
the personalization and tend to return the same results for people
with far different needs. Being an effective tool to address this
problem, recommender systems have become a promising way to filter
out the irrelevant information and recommend potentially interesting
items to the target user by analyzing their interests and habits
through their historical behaviors
\cite{Adomavicius05,Herlocker2004,Konstan1997,Huang2004,Liu2009,Liu2008b,Duo2009,Zhou2007b}.
Motivated by its significance in economy and society, the design of
an efficient recommendation algorithm becomes a joint focus of
theoretical physics \cite{Zhang2007a,PNAS}, computer science
\cite{Adomavicius05} and management science \cite{Huang2004}.

Zhang {\it et al.} \cite{Zhang2007a} proposed a new information
framework based on the heat conduction process, namely
heat-conduction-based (HC) recommendation model. HC model supposes
that the objects one user has collected have the recommendation
power to help the target user find potentially relevant objects. If
the target user is replaced by a specific object, HC model is
similar with the collaborative filtering (CF) method, in which the
users rated one target object have the recommendation powers to
identify the potentially interesting users. So far, CF method has
been successfully applied to many online applications and has
becomes one of the most successful technologies for recommender
systems
\cite{Liu2009,Liu2008b,Duo2009,Zhou2007b,PhysicaA2011,Pan2010}. For
example, Herlocker {\it et al.} \cite{2} proposed an algorithmic
framework referring to the user similarity. Luo {\it et al.}
\cite{5} introduced the concepts of local and global user similarity
based on surprisal-based vector similarity and the concept of
maximum distance in graph theory. Sarwar {\it et al.} \cite{6}
proposed the item-based CF algorithm by comparing different items.
Deshpande {\it et al.} \cite{7} proposed the item-based top-$N$ CF
algorithm, in which items are ranked according to the frequency of
appearing in the set of similar items and the top-$N$ ranked items
are returned. Gao {\it et al.} \cite{8} incorporated the user
ranking information into the computation of item similarity to
improve the performance of item-based CF algorithm. Recently, some
physical dynamics, including random walks
\cite{Newman2006,Zhou2007,Liu2009} and the heat conduction process
\cite{Zhang2007a}, have found their applications in node similarity
measurement. Liu {\it et al.} \cite{Liu2009} used the random walks
to calculate the user similarity and found that the modified CF
algorithm has remarkably higher accuracy. As a benchmark for
comparison, we call it standard CF algorithm (hereinafter, CF only
stands for the collaborative filtering using random-walk-based user
similarity \cite{Liu2009}). By considering the high-order
correlation of the users and objects, Zhou {\it et al.}
\cite{Zhou2007b} and Liu {\it et al.} \cite{Liu2008b} proposed the
ultra accurate algorithms, in which the second-order correlations
are used to delete the redundant information. Besides reliably
accurate recommendations, it is also important for recommender
systems to help most individuals find diverse niche objects. CF
algorithms generate recommendations according to similar users'
suggestions, which prefers to rank the popular objects at the top
positions of recommendation lists, leading to high accuracy but low
diversity.

Random walks have been used to quantify trajectories in a symmetric
ways, namely in- and out-diversity and accessibility
\cite{add1,add2,add3}. In this paper, We argue that the opinions of
small-degree users should be enhanced to generate diverse
recommendations, and present a new directed random-walk-based CF
algorithm, namely NCF algorithm, to investigate the effect of user
similarity direction on recommendations. The numerical results on
the data sets, {\it Netflix} and {\it Movielens}, show that the
accuracy of NCF outperforms the state-of-the-art CF methods with
greatly improved diversity, which suggests that the similarity
direction is an important factor for user-based information
filtering.

\begin{figure}
\center\scalebox{0.3}[0.3]{\includegraphics{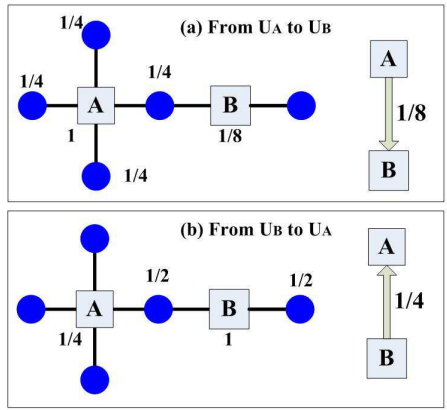}}
\caption{(Color online) Illustration of random-walk-based user
similarity calculation, which has been used to measure user or
object similarity in personalized recommendation. (a) The
possibility walking from user $u_A$ to user $u_B$ is used to measure
the directed user similarity $s_{BA}=1/8$. (b) Similarity from user
$u_B$ to user $u_A$ is $s_{AB}=1/4$. The degrees of user $u_A$ and
$u_B$ are $k_{u_A}=4$ and $k_{u_B}=2$, and one has
$\frac{k_{u_A}}{k_{u_B}}=\frac{0.25}{0.125}$.}\label{Fig.1}
\end{figure}

%\section{Network-based Recommendation Algorithm}

\section{Bipartite Network and heat-conduction-based model}
A recommendation system consists of a set of nodes, object nodes and
connections between two nodes corresponding to an object voted or
collected by a user, which could be represented by a bipartite
network $G(U,O,E)$. We denote the object set as $O=\{ o_1, o_2,
\dots, o_m \}$, the user set as $U=\{ u_1, u_2, \dots, u_n \}$ and
the connection set as $E=\{e_1,e_2,\cdots,e_q\}$. The bipartite
network can then be represented by an adjacent matrix $A=\{a_{\alpha
j}\}\in R^{m,n}$, where $a_{\alpha j}=1$ if $o_\alpha$ is collected
by $u_j$, and $a_{\alpha j}=0$ otherwise.

The final aim of recommender systems is to identify a given users'
interesting objects and generate a ranking list of the target user's
uncollected objects according to the predicted scores. HC model
supposes the neighbor nodes of one target node as the heat sources
with temperature 1, while the remaining nodes are of temperature 0.
According to the thermal equilibrium \cite{Zhang2007a}, the
temperature of the remaining nodes are set as the predicted scores,
which could be calculated by solving the equation ${\bf
W}^hH=\mathbf{f}$, where $\mathbf{f}$ is the flux vector
\cite{Zhang2007a}. The standard HC model firstly constructs the
propagator matrix ${\bf W}^h$, where the element $w_{\alpha\beta}$
denotes the conduction rate from object $o_\beta$ to $o_\alpha$, and
set the temperatures of target node's neighbors as 1, then the heat
will diffuse between heat sources and other nodes. Finally, the
temperatures of uncollected objects are considered as recommendation
scores.

The general framework of the item-based HC model is as follows: (i)
construct the weighted object network (i.e. determine the matrix
${\bf W}$) from the known user-object relations; (ii) determine the
initial resource vector $\mathbf{f}$ for each user; (iii) get the
final resource distribution via
\begin{equation}
\mathbf{f}'={\bf W}\mathbf{f},
\end{equation}
(iv) recommend those uncollected objects with highest final scores.
Note that the initial configuration $\mathbf{f}$ is determined by
the user's personal information, thus for different users, the
initial configuration is different. For a given object $o_\alpha$,
the $i$th element of $\mathbf{f}^\alpha$ should be zero if
$a_{\alpha i}=0$. That is to say, one should not put any
recommendation power (i.e. resource) onto an unrated user. The
simplest case is to set a uniformly initial configuration as
\begin{equation}
f^\alpha_i=a_{\alpha i}.
\end{equation}
Under this configuration, all the users rated object $o_\alpha$ have
the same recommendation power.

The traditional HC model is implemented based on the object
similarity. If the number of users is much smaller than the one of
objects, we could apply the similar idea based on the user
similarity, namely user-based HC model. From the definition, one can
find that user-based HC model is equivalent to the CF algorithm,
therefore, the user similarity analysis of the CF algorithm could
also bring deep insight into the HC model. Zhou {\it et al.}
\cite{Zhou2007} used the random walk process to calculate the node
similarity of bipartite networks and proposed the network-based
inference (NBI) recommendation algorithm \cite{ZhouT2008}. In Ref.
\cite{PNAS}, NBI algorithm was also referred as Probs algorithm. Liu
{\it et al.} \cite{Liu2009} embedded the random walk process into
the CF algorithm to calculate the user similarity and found that the
algorithmic accuracy is greatly increased. In Ref. \cite{Liu2009}, a
certain amount of resource is associated with each user, and the
weight $s_{ij}$ represents the proportion of the resource $u_j$
would like to distribute to $u_i$. This process works by using the
random walk process on user-object bipartite networks, where each
user distributes his/her initial resource equally to all the objects
he/she has collected, and then each object sends back what it has
received to all the users who collect it. The weight $s_{ij}$
representing the amount of initial resource $u_j$ evenly transferred
to $u_i$ can be defined as
\begin{equation}\label{equation2}
s_{ij}=\frac{1}{k_{u_j}}\sum_{l=1}^m\frac{a_{li}a_{lj}}{k_{o_l}},
\end{equation}
where $k_{u_i}$ and $k_{o_l}$ indicate the degrees of user $u_i$ and
object $o_l$.

In the random walk process, the user similarity from user $u_j$ to
$u_i$, $s_{ij}$, is determined by the degrees of commonly rated
objects and user $u_j$'s degree $k_{u_j}$. It is unlikely these
quantities are exactly the same for each pair of users, therefore,
$s_{ij}\neq s_{ji}$ in most cases.

\begin{figure}
\center\scalebox{0.8}[0.8]{\includegraphics{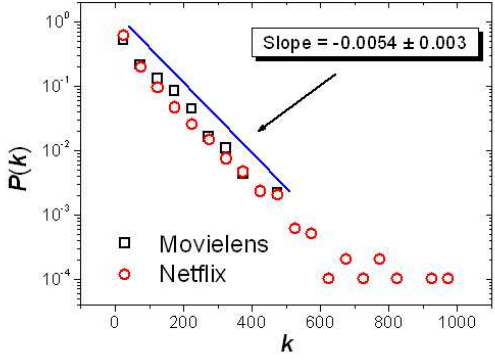}}
\caption{(Color online) User degree distributions for Movielens and
Netflix data sets, which approximately have exponential forms
$P(k)\sim {\rm exp}(-0.0054 k\pm 0.003)$.}\label{Fig1}
\end{figure}

\section{Effect of User Similarity Direction on CF algorithm}
According to the random-walk-based user similarity calculation (See
Fig.~\ref{Fig.1}), the user similarity is also calculated by random
walks in a symmetric way. In CF algorithm, the system should
identify the target user's interesting objects with the help of his
neighbors' historical selections or collections. Therefore, after we
obtain the user similarity matrix, the similarities from neighbors
to the target user are used to evaluate the predicted scores.
According to Eq.~\ref{equation2}, to one pair of users $u_i$ and
$u_j$, their similarities could be written as
\begin{equation}
\begin{array}{rcl}
s_{ij}&=&\frac{1}{k_{u_j}}\sum_{l=1}^m\frac{a_{li}a_{lj}}{k_{o_l}},\ {{\rm From}\ u_j \ {\rm to} \ u_i};\\[10pt]
s_{ji}&=&\frac{1}{k_{u_i}}\sum_{l=1}^m\frac{a_{li}a_{lj}}{k_{o_l}},\
{{\rm From }\ u_i \ {\rm to} \ u_j}.
\end{array}
\end{equation}
Therefore, one has
\begin{equation}\label{equation8}
\frac{s_{ij}}{s_{ji}}=\frac{k_{u_i}}{k_{u_j}}.
\end{equation}
If $k_{u_i}$$>$$k_{u_j}$, then $s_{ij}$$>$$s_{ji}$ and vice versa.
For Movielens and Netflix data sets, the exponential forms of user
degree distributions indicate that most users' degree are very small
(See Fig.~\ref{Fig1}), which means that large-degree users would
frequently be identified as small-degree users' friends. As a
consequence, most users' recommendation lists would be similar.

In order to investigate the effect of user similarity direction on
CF algorithms, we introduce a new user similarity direction
generated by random walks from neighbor set $U_n$ to the target user
to measure user similarities and calculate the predicted score
$v_{i\alpha}$. The NCF algorithm could be described as follows:
Firstly, calculate directed user similarities according to
Eq.~\ref{equation2}, then calculate the predicted scores for target
user $u_i$'s uncollected objects by
\begin{equation}\label{equation3}
v_{i\alpha}=\frac{\sum_{j=1}^n s_{ij}^\beta a_{\alpha
j}}{\sum_{ij}^n s_{ij}^{\beta}},
\end{equation}
where $\beta$ is a tunable parameter to investigate the effect of
similarity strength on the algorithmic performance, and $s_{ij}$ is
the similarity from user $u_j$ to $u_i$. When $\beta=1$, all the
user similarities are given the same weight; when $\beta>1$, the
preferences of users with larger similarities are strengthened; when
$\beta<1$, the ones with smaller similarities are strengthened. The
numerical results indicate that changing the user similarity
direction could not only accurately identify user's interests, but
also increase the algorithmic capability of finding niche objects.

\begin{table}
\caption{Basic statistics of the tested data sets.}
\begin{center}
\begin{tabular} {ccccc}
  \hline \hline
   Data Sets      &  Users   & Objects & Links & Sparsity \\ \hline
   MovieLens      &  1,574    & 943     & 82,580 &  $5.56\times 10^{-2}$\\
   Netflix        &  10,000  & 6,000  & 701,947    & $1.17\times 10^{-2}$ \\
 %  Delicious      &  10,000  & 232,657 & 1,233,997 & $5.30\times 10^{-4}$\\
   \hline \hline
    \end{tabular}
\end{center}
\end{table}

\section{Maximal-Similarity-based CF algorithm}
The algorithmic performance may be affected by the user similarity
direction, and may also be determined by the properties of the data
set. In other words, although the algorithmic performance of the NCF
algorithm is much better than the one of the CF algorithm, it may
only happen on specific data sets whose similarities from neighbors
to the target user are more effective than the ones in the opposite
direction. In order to make it clear, we present a
maximal-similarity-based CF (MCF) algorithm to investigate the
influence of similarity magnitude, in which the predicted score from
user $u_i$ to the uncollected object $o_\alpha$, $v_{i\alpha}$, is
given by
\begin{equation}\label{equation7}
v_{i\alpha}=\frac{\sum_{j=1}^n s_{\rm max}^\beta a_{\alpha
j}}{\sum_{j=1}^n s_{\rm max}^\beta},
\end{equation}
where $s_{\rm max}$ is defined as the larger similarity between user
$u_i$ and $u_j$
\begin{equation}
s_{\rm max}=\max\{s_{ij}, s_{ji}\}.
%s_{\rm max}=\{\begin{array}{lr} s_{ij} & {\rm if} \ s_{ij}\geq s_{ji} \\
%s_{ji} & {\rm if} \ s_{ij}<s_{ji} \\
%\end{array}.
\end{equation}
For example, set the similarity from $u_i$ to $u_j$ is
$s_{ji}=0.01$, while the one from $u_j$ to $u_i$ is $s_{ij}=0.9$.
When recommending objects to $u_i$, the larger similarity 0.9 is
used regardless of the similarity direction.

\begin{table}
\caption{Performances of NCF, CF and MCF algorithms for Netflix and
Movielens data sets according to each of five metrics. The
popularity $\langle k\rangle$, diversity $S$, precision $P$ and
recall $R$ corresponding to $L=10$.}
\begin{center}
\begin{tabular}{cc|ccccc}
\hline\hline
%Algorithms & & &  Netflix & & & & & Movielens & & \\ \hline
                &       & $\langle r\rangle$ & $\langle k\rangle$ & $S$ & $P$ & $R$
                \\ \hline
                &   NCF & {\bf 0.0450} & {\bf 2506} & {\bf 0.8236} & {\bf 0.0967} & {\bf 0.1640}\\
  {\it Netflix} &   CF  & 0.0497 & 2813 & 0.7001 & 0.0917 & 0.1365\\
                &   MCF & 0.0477 & 2758 & 0.7378 & 0.0954 & 0.1374\\
\hline
                &   NCF & {\bf 0.0864} & {\bf 237} & {\bf 0.8929} & {\bf 0.1502} & {\bf 0.2037}\\
 {\it Movielens}&   CF  & 0.1037 & 275 & 0.8435 & 0.1497 & 0.2010\\
                &   MCF & 0.0970 & 271 & 0.8434 &0.1459 & 0.1936
                \\ \hline\hline
\end{tabular}
\end{center}
\end{table}

\begin{table}
\begin{center}
\caption{Algorithmic performances for Movielens data when $p=0.9$,
including the average ranking score $\langle r\rangle$, diversity
$S$ and popularity $\langle k\rangle$ corresponding to length of
recommendation list $L=50$. GRM is a global ranking method; CF is
the collaborative filtering algorithm based on random walks
\cite{Liu2009}; Heter-CF is a modified CF algorithm, in which the
user similarity is defined based on the mass diffusion process and
the second-order similarity is involved ($\beta_{\rm opt}=-0.82$)
\cite{Liu2008b}.\; CB-CF refers to the CF algorithm on weighted
bipartite network \cite{PhysicaA2011}; Hybrid is an abbreviation of
the hybrid algorithm proposed in Ref. \cite{PNAS}. Each number is
obtained by a averaging over five ten runs with independently random
division of training set and probe set.}
\begin{tabular}{cccc}
\hline\hline Algorithms & {\rm Ranking score} & {\rm Popularity} &
{\rm
Hamming Distance}\\
\hline
GRM  & 0.1390 & 259 & 0.398\\
CF & 0.1063 & 229 & 0.750\\
Heter-CF & 0.0877 & 175 & 0.826\\
CB-CF & 0.0914 & 148 & 0.763 \\
Hybrid & 0.0850 & 167 & 0.821 \\
NCF & 0.0864 & 178 & 0.801\\
\hline\hline
\end{tabular}
\end{center}
\end{table}

\section{Simulation Results}

\subsection{Data Description}
In this paper, we base our simulation results on two data sets. The
{\it Movielens} \footnote{http://www.Movielens.com} data set
consists of 100,000 ratings from 943 users on 1,574 movies(objects)
and rating scale from 1 (i.e., worst) to 5 (i.e., best). The {\it
Netflix} data set \cite{add5} is a random sample of the whole
records of user activities in Netflix.com, which consists of 6,000
movies and 10,000 users and 824,802 ratings. The users of Netflix
also vote movies by discrete ratings from one to five. Here we apply
a coarse graining method: a movie is considered to be collected by a
user only if the rating is larger than 2. In this way, the Movielens
data has 82,580 edges, and the Netflix data has 701,947 edges (See
Table I for basic statistics). As an online movie recommendation web
site, Movielens invites users to rate movies and, in return, makes
personalized recommendations and predictions for movies the user has
not already rated, under-contribution is common. Unlike Netflix web
site, MovieLens does not have any DVD rental service. The data set
$E$ is randomly divided into two parts $E=E^T \cup E^P$, where the
training set $E^T$ is treated as known information, containing $p$
percent of the data, and the remaining $1-p$ part is set as the
probe set $E^P$, whose information is not allowed to be used for
prediction.

\subsection{Metrics}

\subsubsection{Average ranking score} Accuracy is one of the most
important metric to evaluate the recommendation algorithmic
performance. An accurate method will put preferable objects in
higher places. Here we use {\it average ranking score}
\cite{Zhou2007} to measure the accuracy of the algorithm. For an
arbitrary user $u_i$, if the object $o_\alpha$ is not collected by
user $u_i$, while the entry $u_i$-$o_\alpha$ is in the probe set, we
use the rank of $o_\alpha$ in the recommendation list to evaluate
accuracy. For example, if there are 8 uncollected objects for user
$u_i$, and object $o_\alpha$ is ordered at the 3rd place, we say the
position of $o_\alpha$ is 3/8, denoted by $r_{i\alpha}=0.375$. Since
the probe entries are actually collected by users, a good algorithm
is expected to give high recommendations to them, leading to a small
$r_{i\alpha}$. Therefore, the mean value of the positions, averaged
over all the entries in the probe set, can be used to evaluate the
algorithmic accuracy.
\begin{equation}
\langle r\rangle=\frac{1}{n}\sum_{i=1}^n\Big(\frac{\sum_{(u_i,
o_\alpha)\in E^p}r_{i\alpha}}{q-k_{u_i}}\Big),
\end{equation}
where $E^p$ is the edge set existing in the probe set and $q$ is the
number of objects in the probe set. The smaller the average ranking
score, the higher the algorithmic accuracy, and vice verse.

\subsubsection{Precision and Recall} Since users usually only
consider the top part of the recommendation list, a more practical
metric is to consider the number of users' hidden links ranked in
the top-$L$ places. We adopt another accuracy measure called {\it
precision}. For a target user, the precision is defined as the ratio
between relevant objects (namely the objects collected by $u_i$ in
the probe set) and the length $L$. Averaging the individual
precisions over all users, we obtain the mean value $P(L)$ of the
algorithm on one data set,
\begin{equation}
P(L) = \frac{1}{n}\sum_i\frac{d_i(L)}{L},
\end{equation}
where $d_i(L)$ indicates the number of relevant objects existing in
the top-$L$ places of the recommendation list.  A larger precision
corresponds to a better performance. {\it Recall} is defined as the
ratio between the number of objects existing in the top-$L$ places
of the recommendation list and the total number of collected objects
$C_i$ in the probe set. Averaging over the individual recalls, we
obtain the mean recall $R(L)$, which could be defined as
\begin{equation}\label{equation6}
R(L)=\frac{1}{n}\sum_i\frac{d_i(L)}{C_i}.
\end{equation}
The larger recall corresponds to the better performance.

\subsubsection{Diversity}
The analysis results on Facebook data set showed that, besides the
common interests, users of online social networks also have their
specific tastes and interests \cite{JPPNAS}, leading to diverse
selection behaviors. Liu {\it et al.} \cite{PRE2011} found that
users' tastes on Movielens and Netflix data could also be divided
into two categories: {\it common interests} and {\it specific
interests}. Therefore, besides accuracy, the diversity of all
recommendation lists is taken into account to evaluate the
algorithmic performance. In general, most of the users would not
show negative altitude to popular objects, therefore, ranking
popular objects at the top part of recommendation lists would
generate higher accuracy. However, personalized recommendation
algorithms should not only present accurate prediction but also
generate different recommendations to different users according to
their specific tastes or habits. The diversity can be quantified by
the average Hamming distance,
\begin{equation}\label{equation4}
S= 1-\langle Q_{ij}(L)\rangle/L,
\end{equation} where
$L$ is the length of the recommendation list and $Q_{ij}$ is the
number of overlapped objects in $u_i$'s and $u_j$'s recommendation
lists. The largest $S=1$ indicates recommendations to all users are
completely different, in other words, the system has the highest
diversity. While the smallest $S=0$ means all of recommendations are
exactly the same.

\subsubsection{Popularity} An accurate and diverse recommender system
is expected to help users find the niche or unpopular objects which
are hard for them to identify. The metric {\it popularity} is
introduced to quantify the capacity of an algorithm to generate
unexpected recommendation lists, which are defined as the average
collected times over all recommended objects
\begin{equation}
\langle k\rangle =
\frac{1}{n}\sum_i\Big(\frac{1}{L}\sum_{o_\alpha\in
O^L_i}k_{o_\alpha}\Big),
\end{equation}
where $O^L_i$ is $u_i$'s recommendation list with length $L$. A
smaller average degree $\langle k\rangle$, corresponding to less
popular objects, are preferred since those lower-degree objects are
hard to be found by users themselves.

\begin{center}
\begin{figure}
\center\scalebox{0.8}[0.7]{\includegraphics{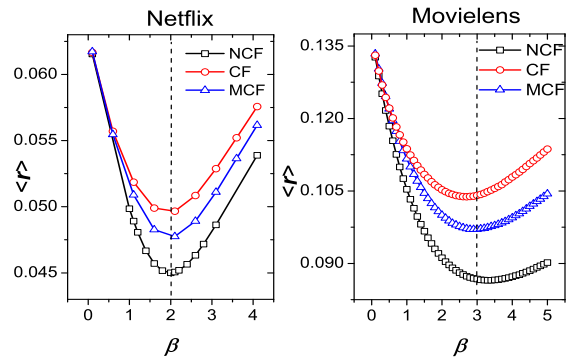}}
\caption{(Color Online) The average ranking score $\langle r\rangle$
vs. $\beta$ for NCF, CF and MCF algorithms. The optimal $\beta_{\rm
opt}$ of NCF for Movielens data set, corresponding to the minimal
$\langle r\rangle= 0.086$, is $\beta_{\rm opt}=3.3$, the one for
Netflix data set is $\beta_{\rm opt}=2.0$ corresponding to $\langle
r\rangle=0.0450$. When $\beta=1$, the algorithm degenerates to the
accuracy of the CF algorithm based on the new user similarity
direction. All the data points are averaged over ten independent
runs with different data-set divisions.}\label{Fig2}
\end{figure}
\end{center}

%\begin{center}
\begin{figure*}
\scalebox{0.85}[0.85]{\includegraphics{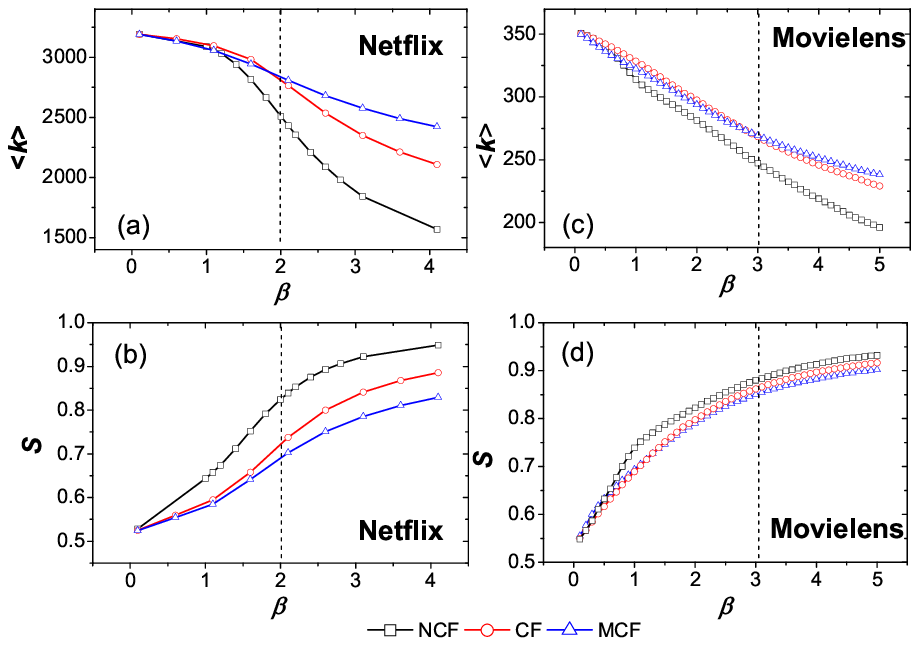}}
\scalebox{0.76}[0.76]{\includegraphics{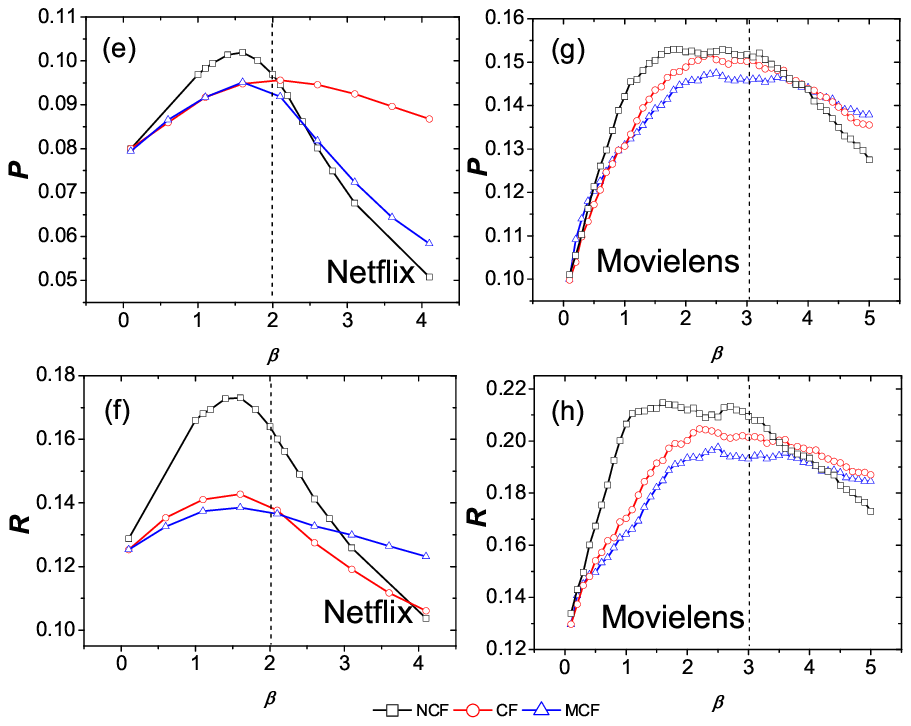}}\caption{(Color
online) Performances of the NCF, CF and MCF algorithms for MovieLens
and Netflix data sets when recommendation lists equal to $L=10$.
(a)-(d) show average object degrees $\langle k\rangle$ vs. $\beta$
and diversity $S$ vs. $\beta$. At the optimal cases, both popularity
$\langle k\rangle$ and diversity $S$ of NCF are much better than the
ones of CF and MCF algorithms. (e)-(h) Precision $P$ and recall $R$
vs. $\beta$ for Netflix and Movielens data sets. One can find that
both of $P$ and $R$ of NCF for Movielens are larger than the ones of
CF and MCF algorithms, while the precision $P$ for Netflix is close
to the one of CF algorithm and the recall $R$ is much better than
the results of CF algorithm. All the data points are averaged over
ten independent runs with different divisions of training-probe
sets.}\label{Fig3}
\end{figure*}
%\end{center}

\subsection{Simulation Results}
We summarize the results for NCF, CF and MCF algorithms, as well as
the metrics for Movielens and Netflix data sets in Table II.
Clearly, NCF outperforms the classical CF and MCF algorithms over
all five metrics, including average ranking score $\langle
r\rangle$, diversity $S$, popularity $\langle k\rangle$, precision
$P$ and recall $R$. Table III gives the comparisons among different
algorithms for $p=0.9$. The so-called optimal parameters are subject
to the lowest average ranking score $\langle r\rangle$. The metrics,
including average ranking score $\langle r\rangle$, diversity $S$
and popularity $\langle k\rangle$, are obtained at the optimal
parameters. From which one can see that the accuracy of NCF is close
to the result of hybrid algorithm \cite{PNAS} and outperforms the
state-of-the-art CF algorithms using the second-order correlation
information \cite{Liu2008b}. Among all algorithms, Heter-CF
algorithm gives the highest diversity, while CB-CF algorithm
generates the lowest popularity. Comparing with these two
outstanding algorithms, NCF can reach or closely approach the best
diversity without taking into account high-order correlation, and
provide more accurate recommendation results.

Figure 3 reports the algorithmic accuracy as a function of $\beta$.
In our algorithm, the curve (the red circle) has a clear minimum
around $\beta_{\rm opt}=3.3$ for Movielens and $\beta_{\rm opt}=2.0$
for Netflix. Comparing with CF algorithms whose user similarities
are defined from large-degree to small-degree users, the average
ranking score $\langle r\rangle$ of NCF is reduced from 0.497 to
0.045 for Netflix and from 0.1037 to 0.0864 for Movielens data set,
reduced by 9.9\% and 16.68\% respectively at the optimal values.
Comparing with the MCF algorithm, the performance of NCF is also
better. Subject to the accuracy, the reason why NCF outperforms CF
and MCF indeed lies in the direction effect but not the data effect,
and the results also indicate that giving more recommendation power
to the small-degree users could enhance the accuracy and diversity
simultaneously.

The Hamming distance $S$ is introduced to measure the algorithmic
performance to present personalized recommendation lists. The
average object degree $\langle k\rangle$ is used to evaluate the
ability that an algorithm gives a novel recommendation. Figure 4
(a)-(d) show $\langle k\rangle$ and $S$ as a function of $\beta$
when recommendation list length $L=10$ respectively. For the
Movielens data set, at the optimal point $\beta_{\rm opt}=3.3$, the
popularity $\langle k\rangle = 237$, which is reduced by 13.8\%, and
the diversity $S=0.8929$ is improved by 5.9\% comparing with the
ones of CF at its optimal value. When the list length $L=10$, the
popularity $\langle k\rangle$ and diversity $S$ of NCF are reduced
by 10.9\% and 17.64\% for Netflix data set. From which one can find
that the NCF algorithm using the new directed random walks has the
capability to find the niche objects, leading to diverse
recommendations.

In general, NCF outperforms CF as well as MCF in terms of the
accuracy $\langle r\rangle$, diversity $S$ and popularity $\langle
k\rangle$. However, in reality, users only care about the top part
of the recommendation list. From Fig. 4 (e)-(h), one can find that,
comparing with the results of CF and MCF algorithms, the precision
$P$ and recall $R$ of NCF are also very good. When $L=10$ with the
optimal parameter corresponding to the lowest ranking score, the
precision $P$ is approximately improved 3.0\% and 5.5\%, and the
recall $R$ is roughly enhanced by 5.2\% and 20.15\% for Movielens
and Netflix data sets respectively.

\begin{center}
\begin{figure}
\scalebox{0.9}[0.9]{\includegraphics{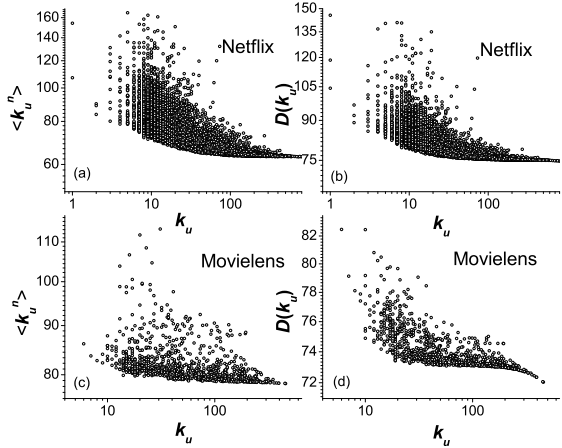}} \caption{(a)-(b)
The average degree of the target user's neighbors $\langle
k_u^n\rangle$ and the corresponding deviation $D(k_u)$ vs. target
user degree $k_u$ for Netflix data set. (c)-(d) show the results for
Movielens. To small-degree users, both their neighbors' average
degree $\langle k_u^n\rangle$ and the deviation $D$ are very large.
As $k_u$ increases, $\langle k_u^n\rangle$ and $D(k_u)$ would
decease correspondingly.}\label{Fig6}
\end{figure}
\end{center}

%\subsection{Degree correlation analysis of user pairs}
Since the similarities generated by the random walk process from
small-degree to large-degree users are larger than the ones from the
opposite direction, the simulation results indicate that enhancing
the small-degree users' recommendation powers increases the
prediction accuracy, and helps users find niche objects, leading to
more diverse recommendations. Figure \ref{Fig6} investigates the
correlation between the target user degree $k_u$ and its neighbors'
average degree $\langle k_u^n\rangle$ as well as deviation $D(k_u)$,
where the target user's neighbors $U_n$ are defined as the users who
have at least one common rated object with the target user, which
could be obtained from the adjacent matrix $A$. Denoting the user
correlation matrix as $C^{\rm user}$, we have $C^{\rm user}=AA^T$.
The element $C^{\rm user}_{ij}$ means the number of common rated
objects between user $u_i$ and $u_j$. Given a matrix
$T=\{t_{ij}\}\in R^{n,n}$, with $t_{ij}=1$ if $C^{\rm user}_{ij}>
0$, and $t_{ij}=0$ if $C^{\rm user}_{ij}=0$. The number of
correlated neighbors $k^c_u$ for a target user $u$ could be given as
$k^c_u=\sum_{j=1}^n t_{uj}$, then average degree $\langle
k_u^n\rangle$ of correlated neighbors $U_n$ is defined by
\begin{equation}
\langle k_u^n\rangle = \frac{1}{k^c_u}\sum_{j=1}^{n}t_{uj}k_j.
\end{equation}
The deviation $D(k_u)$ could be given as
\begin{equation}
D(k_u) = \sqrt{\frac{1}{k^c_u}\sum_{j=1}^n\Big(t_{uj}k_j-\langle
k_u^n\rangle\Big)^2}.
\end{equation}
Figure~\ref{Fig6} shows that when $k_u$ is very small, both
neighbors' average degree $\langle k_u^n\rangle$ and deviation
$D(k_u)$ are very large, which means that for Movielens and Netflix
data sets, the small-degree users would like to commonly rate
objects with small-degree users and large-degree users. As $k_u$
increases, both $\langle k_u^n\rangle$ and $D(k_u)$ would decrease
correspondingly, which means that the large-degree users only
commonly rate objects with small-degree users. According to our
previous analysis, if the user similarities from neighbors to the
target user are enhanced, the effects of the small-degree users
would be emphasized to match both large-degree and small-degree
users' common and specific interests, which is the reason why
directed random-walk-based user similarity is effective.

\section{Effects of data sparsity}

We investigate the effects of the data sparsity on the performance.
Since we focus on the similarity direction effect on the CF
algorithm, we choose the classical CF algorithm for comparison. In
the simulation work, we select $pE$ edges as training set, and set
the rest $(1-p)E$ edges as probe set. Lower $p$ means less
information is used to generate the recommendations. The numerical
results for Movielens are shown in Fig.~6. Each point of the
histogram is obtained with the optimal parameter subject to the
lower ranking score. The improvement function $f(\langle r\rangle)$
of the present algorithm is defined as
\begin{equation}
f(\langle r\rangle)=\frac{\langle r\rangle_{\rm CF}-\langle
r\rangle_{\rm NCF}}{\langle r\rangle_{\rm CF}}.
\end{equation}
For the popularity $\langle k\rangle$ and Hamming distance $S$, the
improvement functions are defined as
\begin{equation}
\begin{array}{c}
f(\langle k\rangle)=\frac{\langle k\rangle_{\rm CF}-\langle
k\rangle_{\rm NCF}}{\langle k\rangle_{\rm CF}},\\[12pt]
f(S)=\frac{S_{\rm NCF}-S_{\rm CF}}{S_{\rm CF}}.
\end{array}
\end{equation}
Figure 6 shows that the improvement of average ranking score
$\langle r\rangle$ decreases as the size of the training set
decreases, which may be come from the fact that the number of
neighbors would decrease and less information could be used to
predict the target user's interests. We also found NCF performs much
better than CF for denser datasets. The improvement of diversity
$f(S)$ decreases with the increasing size of training set, and
$f(\langle k\rangle)$ increases with more denser training set, which
indicate that, generally speaking, users prefer to select popular
objects as they give more ratings.

\begin{center}
\begin{figure}
\center\scalebox{0.7}[0.7]{\includegraphics{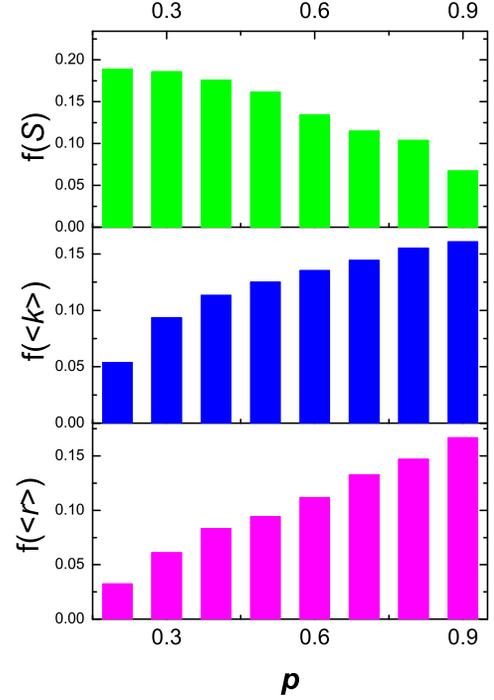}}
\caption{Improvements of the average ranking score $\langle
r\rangle$, average objects degree $\langle k\rangle$ and diversity
$S$ to different sparsity of the training set for Movielens data
set. All the data points are averaged over ten independent runs with
different data set divisions.}\label{Fig5}
\end{figure}
\end{center}

\section{Conclusion and Discussions}
In this paper, by tuning the random walk direction from neighbors to
the target user to calculate the directed user similarity, we
investigate the physics of directed random walks and their influence
on the information filtering of user-object bipartite networks.
Simulation results indicate that the new CF algorithm using the new
directed random walks outperforms state-of-the-art CF methods in
terms of the accuracy, as well as the Heter-CF algorithm in terms of
diversity simultaneously. Meanwhile NCF has much better capability
to present more accurate and diverse recommendations than CF
algorithms whose user similarities are calculated from the target
user to neighbors. The accuracies of NCF algorithms are close to the
results of the hybrid algorithm \cite{PNAS}, and the diversities are
also increased dramatically.

CF algorithms are one of the most successful information filtering
algorithm and has been extensively used on lots of web sites, such
as Netflix and Amazon. HC model also has been successfully used for
information filtering. If we suppose the users rated one specific
objects are the heat source, CF algorithms are equivalent to the
user-based HC model. Although random walks have been used to improve
the user or object similarity measurement \cite{PNAS,Liu2009}, the
reason why directed similarity could enhance the information
filtering performance is missing. Since the idea of the CF algorithm
is combing neighbors' opinions of the target user to predict his
interests or habits, we always suppose that the CF algorithm would
like to converge users' interests and present popular objects. But
the analysis in this paper indicates that if small-degree users'
recommendation powers are increased CF algorithm also could solve
the accuracy-diversity dilemma. In most of the online social
systems, the number of small-degree users is always much larger than
large-degree ones. According to the random-walk-based user
similarity definition, we know that similarities from small-degree
users are always larger than the reversed ones. Therefore, in the CF
algorithm, the opinions of the large-degree users would be
recommended to most of small-degree users, leading to lower
diversity. By tuning the similarities from neighbors to the target
user, we could emphasize the recommendation powers of small-degree
users and enhance the accuracy and diversity simultaneously, which
indicates that the similarity direction is an important factor for
information filtering. Although the idea of this paper is simple,
the remarkable simulation results indicate that, to generate
accurate and diverse recommendations, we only need to change the
direction without changing the framework of the existing CF systems.

The directed random walk process presented in this paper indeed has
been defined as a local index of similarity in link prediction
\cite{Lv2,EPJB2009}, community detection \cite{Liu2010P} and so on.
Meanwhile, a number of similarities, based on the global structural
information, have been used for information filtering, such as the
transferring similarity \cite{Duo2009} and the PageRank index
\cite{Brin1998}, communicability \cite{add4} and so on. Although the
calculation of such measures is of high complexity, it's very
important to the effects of directed random walks on these measures.
The hybrid algorithm \cite{PNAS} is also a kind of item-based CF
algorithm where the item similarity is measured by combining the
random walk and heat conduction processes together. L\"{u} {\it et
al.} \cite{Linyuan2011} proposed an improved hybrid algorithm by
embedding the preferential diffusion process into hybrid algorithm.
Qiu {\it et al.} \cite{zike2011} proposed an improved method by
introducing an item-oriented function to solve the cold-start
problem. In this paper, we find that the direction of random walks
is very important for information filtering, which may be helpful
for deeply understanding of the applicability of directed
similarity.

We thank Dr Chi Ho Yeung for his helpful suggestions. We acknowledge
{\it GroupLens} Research Group for providing us {\it MovieLens} data
and the Netflix Inc. for {\it Netflix} data. This work is partially
supported by NSFC (10905052, 70901010, 71071098 and 71171136), JGL
is supported by the European Commission FP7 Future and Emerging
Technologies Open Scheme Project ICTeCollective (238597), the
Shanghai Leading Discipline Project (S30501) and Shanghai
Rising-Star Program (11QA1404500).

\end{document}